# SLEEPY CHAUFFEUR DETECTION AND ALERT TECHNIQUES FOR ROAD SAFETY

HIMEL GHOSH[*], SAYAK CHATTERJEE[*], ANTIK GANGULY[*], SHREETAMA KARMAKAR[*], KOUSHIK SARKAR[**]

## ABSTRACT

The most startling of the contemporary problems is the sleepiness of chauffeur which causes lots of car accidents. Prevention of those impending accidents by detecting and alerting the sleepy chauffeur is vital, otherwise that would lead to loss of lives and various traumas along with severe injuries. The slumber or sleep may be caused by huge stress, pressure, relentless work load or alcoholism, for which sleep deprivation occurs and the chauffeur while driving gets drowsy. So far, considerable amount of systems has been developed to detect drowsiness of drivers, most of which mainly depend on image processing algorithms using cameras. Some of them also incorporate artificial intelligence and machine learning based algorithms. This paper presents a review of the existing systems and also proposes an easy and cheap system using sensors and Arduino, capable of detecting sleepiness and generates siren alarm and send alert message to take precautionary measures.

## INTRODUCTION

In this busy modern world, people are so engaged with strenuous jobs and relentless pressure, thereby leading to lack of proper sleep causing sleep deprivation, which increases day-time sleepiness affecting the job performance. The chauffeur (i.e. driver) sleepiness is a major concern nowadays, many road accidents occur every year, because of this. In order to ensure safety of both passengers and pedestrians, the drowsiness of the driver must be detected and systems should be designed to prevent any such accidents.

The driver drowsiness detection system came into existence since the 90's. This system would continuously monitor driver's driving performance without attracting any attention of the driver. The system has a preloaded alarm system that will be triggered when drowsiness is detected.

[*]Student of 3rd Year, Electronics and Communication Engineering Department, Future Institute of Engineering and Management, Kolkata-700150.
[**]Assistant Professor, Electronics and Communication Engineering Department, Future Institute of Engineering and Management, Kolkata-700150.
*Correspondence E-mail Id:* editor@eurekajournals.com

     

The key requirements include low cost, true unobtrusiveness, an acceptably-low false alarm rate, non-disruption of the primary driving task, and a warning strategy that truly sustains driver wakefulness or convinces him/ her to stop for rest. The technology of image processing is implemented to analyze the driver's face and eyes. The detection is carried out by detection in Brain waves, blinking, heart rate, pulse rate, skin electric potential and some physical reactions like changes in inclination of Driver's head, sagging posture, frequency at which eyes close, gripping force on steering wheel [1, 2]. As technology advanced a monitoring system developed which uses colour analysis for finding the lips on the driver's face. Then face of the driver is recognized and the skin colour of the driver is sampled. The eyes and the lips are the largest holes in the region. The pupils can be located by marking the darkest pixels in the eyes. The system uses relative positions of the pupils and eyes to make some calculations regarding the gaze direction. Then with the advancement of communication came real-time online driver monitoring system which uses remotely located charged coupled device (CCD) cameras to capture video images of the driver. The best method of detection however are the methods that require physical contact with the drivers to measure heart rate, pulse rate, brain impulses. Electroencephalograph (EEG) was first developed on which offline monitoring equipment are based. There is also an online version present called "mind switch" in which electrodes are embedded in a headband device which measures brain waves.

Ocular measures are however considered to be the most suitable way of monitoring. Physiological measures of third type include facial expressions, body postures, and head nodding [3, 4]. The collection of the chauffeur's head movement data with IR sensors has been used [5], which is directly analyzed by a user to study and classify between normal and sleepy driving and then they applied to their detection system. The requirement of the third-party user to draw the analysis, poses a problem in terms of time, customizability, accuracy, efficiency and cost. The various ways to measure hypo-vigilance of the chauffeurs has been discussed [6], and details on the hypo-vigilance alerting methods using several warning signals to alert the chauffeur is given. We get an insight into the method used to detect the chauffeur's face and eyes in [7]. It uses face detection algorithm based on symmetry, followed by an eye localization process to locate the center of the eye, then they used the SAD (Sum of Absolute Differences) algorithm to track, after which the state of the chauffeur is determined by measuring PERCLOS (Percentage of Eye Closure). IR Camera has been used to serve the purpose under day and night ambience. But the proposed system could detect the drowsiness state of the chauffeur but did not suggest anything to serve as the alert mechanism to stop any accidents. Moreover, limitations in the calculation of symmetry accounts for lesser accuracy of the system. We saw, light has been shed on the contemporary development in technologies, generally used by companies like Toyota, Ford and others for the detection and alerting purpose [8].

The recent developments in the design and idea of the drowsy driver system uses the microcontroller and several sensors connected to it which will not only detect the sleepiness of the chauffeur but also initiate an alarm and vehicle correction in order to prevent any accident [9], the sensors used are IR Detector (to measure eye blinking rate), accelerometer (to detect unusual head nodding), thermistor (to measure breathing rate), IR LED and Phototransistor (to measure heart rate) and the microcontroller Arduino UNO ATmega328p which will be suitably programmed; this system unlike the others with Image processing algorithms, is proposed to work well in the dim light situations which are the more practical cases during night.

This setup renders a more cost-effective way of implementing a chauffeur drowsiness detection and alert system. A system has been proposed much recently, which uses goggles fitted with IR sensors and controlled by Arduino microcontroller that will detect the sleepiness based on an algorithm that checks if the eyes are closed using a loop and send alert signals using GSM module to the owner of the car if the chauffeur is asleep [10].

This setup is not very effective because the chauffeur might not wear those goggles all the time, and even if sleep is detected, there is no proper mechanism to effectively alert the chauffeur which would stop an accident. The future aspects of the system to detect chauffeur's sleepiness and provide effective alert, lies in the proper study of the physiological parameters including slow heart-rate, EEG patterns and also behavioral parameters such as yawning, eye blinking, head nodding. Use of effective algorithms to detect the face and eyes and applying machine learning techniques such as Support Vector Machines (SVM), Convolutional Neural Networks (CNN) or Hidden Markov Models (HMM) to the obtained analysis of physiological and behavioral parameters thereby developing a fully automatic system capable of detecting chauffeur's sleepiness [11] can be built. More research is ongoing regarding that.

## LITERATURE SURVEY

Knipling et al. [1] proposed a system, which is based on the minute steering movements and the eye closure interval of the driver. It continuously analyses the psycho-physiological signs as well as the driver's performance. A slight change in any of these will trigger the system and will provide an immediate warning signal to the driver.

The system described by Hiroshi Ueno et al. [2] uses image processing which analyses the images of drivers face with the help of a digital camera fitted in front the driver. It gets alerted depending on the time interval for which the driver's eyes are open or closed and provides a non-contact method of alerting the driver. Wahlstrom et al. [3] has used dash-board mounted camera along with Framework for Processing Video (FPV) which works by detecting lips of the driver by colour analysis and then skin colour is sampled to generate face region. Qiang Ji et al. [4] has furnished the use of CCD Cameras with IR illuminators to capture video. By analysis of real-time visual cues, a probabilistic model is developed for the purpose of prediction of drowsiness.

Dongwook Lee et al. [5] focused on the study of driver's head-movement data to detect drowsiness and proclaimed to have 78% accuracy. Arun et al. [6] has focussed on the concept of hypo-vigilance of the driver and reviewed the various ways to alert the driver. Belal ALSHAQAQI et al. [7] describes a system based on visual information measuring PERCLOS with eye closure. Nidhi Sinha et al. [8] has reviewed all the existing systems based on Digital Image Processing, EEG, ECG, etc. It has been furnished by Akshay Bhaskar [9] that his system, monitors blinking rate, unnatural head nodding, heart rate etc. It can alert a drowsy chauffeur and can provide visual notification to other drivers. It proclaims to have 70% accuracy. Gabhane et al. [10] has proposed a vehicle based adaptive driver alert system and driving behaviour application to the company owners.

Ngxande et al. [11] thoroughly investigated the existing techniques of behavioural based drowsiness detection using machine learning algorithms and reviews the techniques based on deep learning concepts, besides concluding Support Vector Machine as the most common process of detection. Suryawanshi et al. [12]

studies the time interval between two successive eye blinks. The camera captures and recognizes the face of the driver and then the eyes are located by computing horizontal averages in that region. The change in intensity of the eyes is calculated to determine whether eyes are open or closed. This is actually based on non-intrusive machine vision. Chisty et al. [13] describes an upgraded version of non-intrusive methods. Circular Hough Transform, FCM, Lab Colour Space, Varying Luminance Conditions, etc are used in this system.

## DESIGN AND WORKING PRINCIPLE

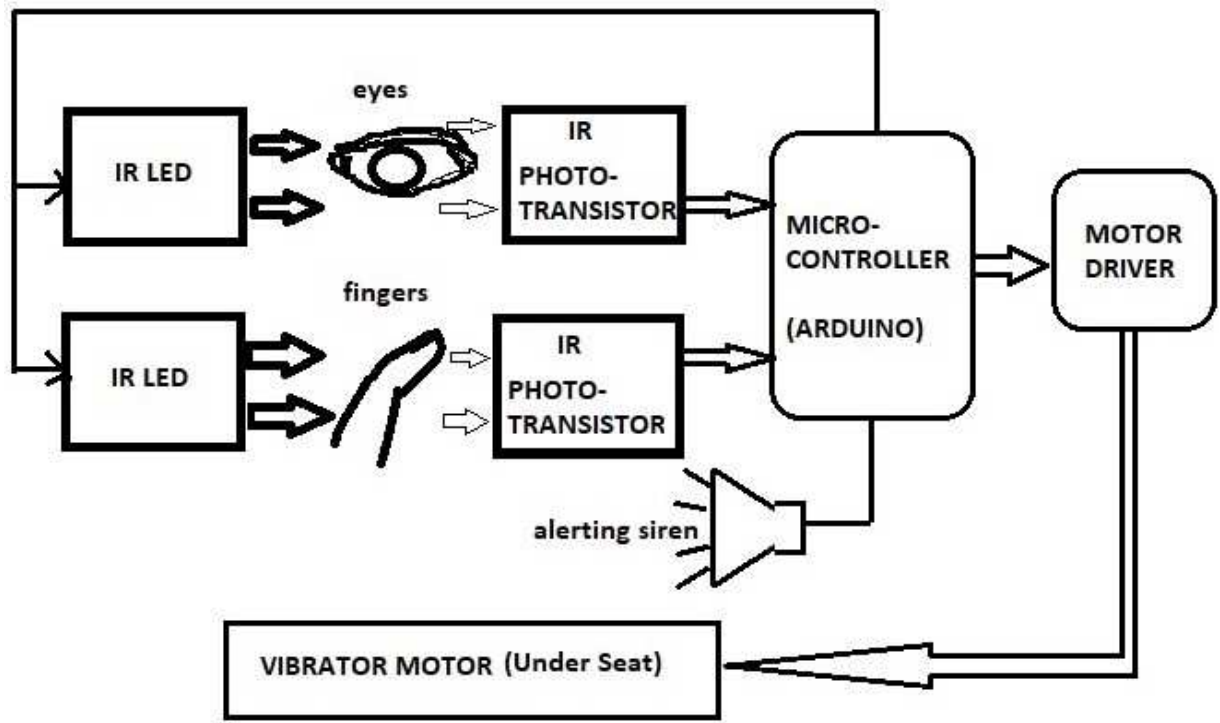

Figure 1: Block diagram

The above Figure 1 shows the design of the driver drowsiness detection as well as the alert mechanism. We have used Infra-red Light Emitting Diodes (IR LED), the IR Phototransistor, Microcontroller (Arduino Uno), alerting Siren (A Piezo Buzzer), Motor-driver, Vibrator motor. The block diagram shows the connection scheme.

The IR transmitter and receiver pair are used not only to detect the eye blinking rate, but also the heart rate. Infrared rays (700nm to 1mm) do not lie in the visible range thereby not affecting or disturbing the eyes. The IR sensors have been used in the indirect incidence mode, where both the emitter and detector are placed side by side. In this system the emitted IR rays gets reflected by the human eyes to the detector photo-transistor which converts the received signals into electrical signals which are sent to the microcontroller. The eyeballs reflect the IR rays but the eyelids do not, thus, when eye is closed, no electrical signal is sent to the Arduino. In this way, the eye blinking rate can be determined at an interval of time by using a counter variable. During the first phase of sleep, heart rate drops which increase the blood flow. In the same manner, the IR sensors are used to calculate heart rate over an interval of time, here the purpose is served by reflection of IR rays from the fingers where more blood delays the signal fed to the microcontroller. If the eye is closed for a time more than the eye blinking interval, then the microcontroller would trigger the buzzer siren which is done by suitable coding. If the situation persists for other few seconds and the heart-rate slows down, then both the buzzer siren and vibrator motor under the seat starts working, in order to bring back the alertness of the chauffeur. To ensure more safety, number of sirens used can be increased. The vibrator motor can also be controlled by the Arduino motor-driver. The IR LEDs are powered by the microcontroller itself. IR sensors are advantageous in darker ambience, i.e. at night instead of image processing by video cameras. Further, a GSM module can also be used with the system, in order to send urgent message to the traffic guard or owner of the car or the road safety cell of the administration, so that others including pedestrians can be alerted [Figure 2]. The system is easy to install and costs less to the user as the components are also easily available at the shops.

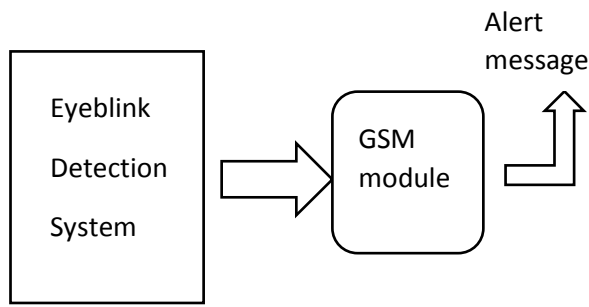

**Figure 2**

The flowchart of the functionality of this system is given below:

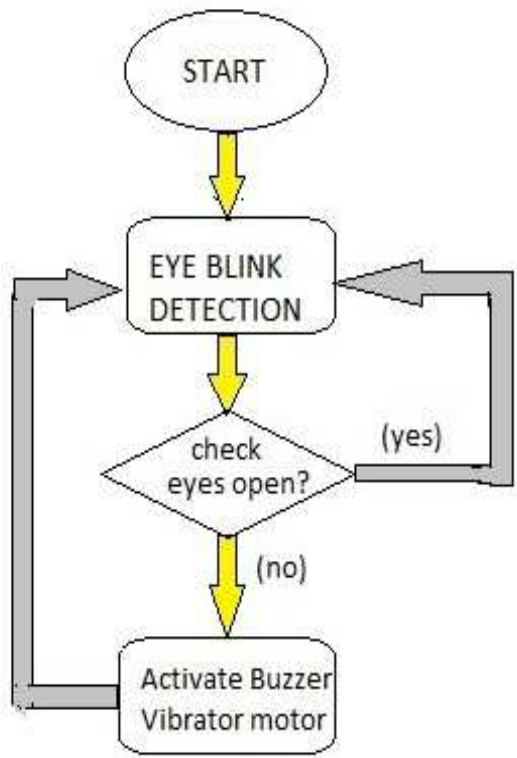

**Figure 3**

The setup will work in this way and the microcontroller will follow the suitable algorithm.

**IR SENSOR**

The IR sensor works at the infrared region (0.75um to 1000um), thus human eyes cannot sense IR rays. The sensor consists of an IR LED as transmitter and IR photodiode as detector. The IR LED illuminates the target and the reflected energy is focused on the IR detector. IR sensors can be used effectively in both day and night [14].

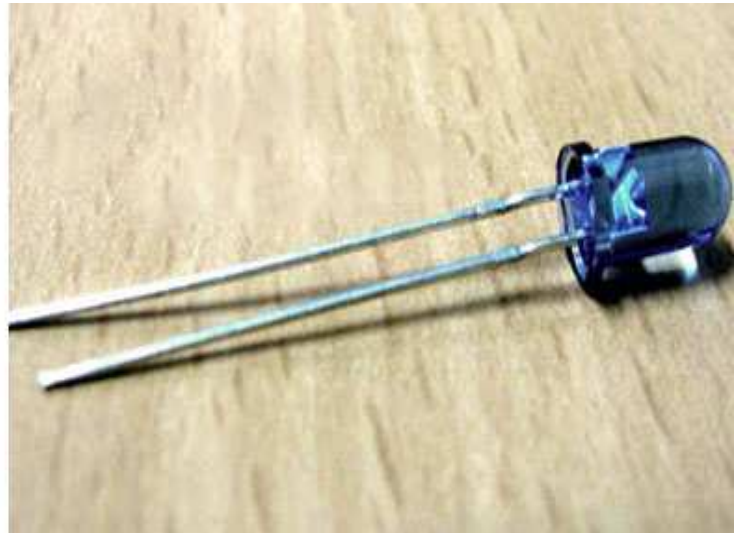

**Figure 4.Jain et al. [14]**

PIEZO BUZZER is based on the principle of piezo-electricity, i.e. electricity is generated at the application of mechanical pressure. This electronic device produces sound used for alerting purposes [15]. Vibrator Motor used is basically the coin vibrators, generally used in cell phones. The GSM module allows mobile communication. It requires a SIM card to connect to the network. It can be used for sending and receiving messages. The GSM module helps to do so by allowing microcontroller to operate it. A MICROCONTROLLER is a small computer on a single integrated circuit [16]. It consists of a CPU, memory element and programmable input-output peripherals. The Arduino Uno is a microcontroller board based on the ATmega328. It has 14 digital input/output pins (of which 6 can be used as PWM outputs), 6 analog inputs, a 16 MHz crystal oscillator, a USB connection, a power jack, an ICSP header, and a reset button [17].

**ADVANTAGES**

The system has easy to implement design and with easily available components. The system is portable and thus, mounting it is easy. The system requires very less power. The system is cheaper compared to others and it works at low light conditions. Most importantly, the system prevents accidents.

**DISADVANTAGES**

This is a very delicate system. This system will not be able to stop the vehicle on its own after the driver gets deeply asleep. It cannot predict the physical and emotional state of the driver. This is system is not so smart. There are certain restrictions of threading, as this system only uses Arduino as microcontroller. The driver can choose not to switch on this system and thus completely avoid its working while driving leading to fatal risks despite the availability of this system.

 

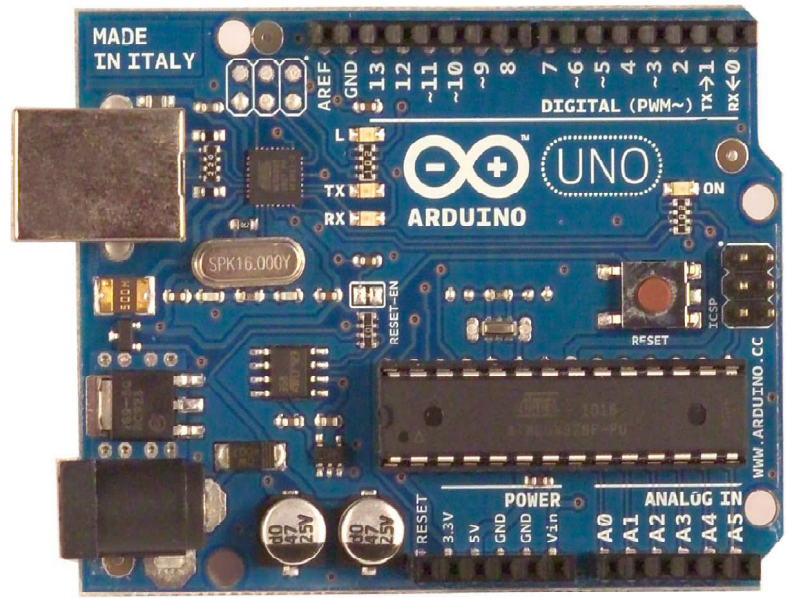

**Figure 5.[Ref. 18]**

## CONCLUSION

In this paper, titled, Sleepy Chauffeur Detection and Alert Techniques for Road Safety, we have gazed through the existing concepts of the drowsiness detection and also reviewed some of the prevalent mechanisms. We also presented a sleek system that stresses not only on the drowsiness detection but also on the alerting techniques so that road safety is given utter priority in order to prevent fatal accidents. We have successfully tested the working of the system in experimental conditions. The system is beneficial in terms of being cheap, portable and easy to implement unlike the prevalent vehicle-based systems.

## FUTURE SCOPES

Detection can include the use of image-processing based algorithms as well as more advanced form of detection procedures can be implemented using Python programming and machine learning algorithms. Apart from eye blink detection, physical state identification, emotional state analysis, blood flow changes in the facial skin can be used for the overall detection of near sleepiness. Even some kind of smart assistance by utilising Artificial Intelligence concepts can be incorporated that would help the driver in coping with the sleepiness and would set the car to safe driving mode automatically or rather stop the car at a nearby parking by predicting the possibility of drowsiness of the chauffeur. Thus, in future, drowsiness would not pose a big problem when such level of advanced systems will be concerned.

## ACKNOWLEDGEMENTS

The authors would like to acknowledge the research and department cell of the ECE department, FIEM, Kolkata, India, for the co-operation and advice extended.